\documentclass[conference]{IEEEtran}
\IEEEoverridecommandlockouts
\usepackage{cite}
\usepackage{amsmath,amssymb,amsfonts}
\usepackage{algorithmic}
\usepackage{graphicx}
\usepackage{textcomp}
\usepackage{xurl}
\usepackage{xcolor}
\usepackage{fancyhdr}
\usepackage{etoolbox}

\makeatother

\def\BibTeX{{\rm B\kern-.05em{\sc i\kern-.025em b}\kern-.08em
    T\kern-.1667em\lower.7ex\hbox{E}\kern-.125emX}}

\begin{document}
\title{Telephony Voice Agent for Banking Services}

\author{\IEEEauthorblockN{Nitya Dhagat}
\IEEEauthorblockA{\textit{Department of Information Technology} \\
\textit{Dharmsinh Desai University}\\
Nadiad, India \\
nityaddhagat@gmail.com \\
}
\and 
\IEEEauthorblockN{Vipul K. Dabhi}
\IEEEauthorblockA{\textit{Department of Information Technology} \\
\textit{Dharmsinh Desai University}\\
Nadiad, India \\
vipuldabhi.it@ddu.ac.in \\}
\and
\IEEEauthorblockN{Harshadkumar B. Prajapati}
\IEEEauthorblockA{\textit{Department of Information Technology} \\
\textit{Dharmsinh Desai University}\\
Nadiad, India \\
prajapatihb.it@ddu.ac.in \\}
\and 
\IEEEauthorblockN{Zankhana J. Barad}
\IEEEauthorblockA{\textit{Department of Information Technology} \\
\textit{Dharmsinh Desai University}\\
Nadiad, India \\
zankhana.it@ddu.ac.in \\}
}

\maketitle

\begin{abstract}
This paper proposes a voice-powered AI-based banking system based on Google Conversational Agent, Dialogflow CX, which provides safe and convenient banking by phone. The system supports essential banking functions such as balance inquiries, transaction history retrieval, card activations, PIN-based authentication of sensitive tasks, smooth live agent handoff for complex and out-of-scope queries, and ensures seamless handover to human agents when required. These tests were performed with high-duration calls, high concurrency, and noisy environments; the system proved to be scalable, responsive, and resilient. All the data used is safely stored in the cloud environment for efficiency and security in real-time voice interactions. A voice-based banking solution that is efficient and easy to use can be provided through this.
\end{abstract}


\section{Introduction}
The banking sector is undergoing rapid digital transformation, driven by evolving customer expectations for real-time, personalized, and accessible financial services \cite{Galileo2024}. This shift has catalyzed the adoption of artificial intelligence (AI), particularly conversational AI, which has progressed from basic, scripted chatbots to advanced virtual agents capable of contextual understanding and complex task execution \cite{Aisera2024} \cite{ClerkChat2024}. Among various interaction modalities, voice interfaces stand out for their naturalness and ease of use, especially in hands-free or accessibility-driven contexts. Modern voice agents aim to surpass the limitations of traditional Interactive Voice Response (IVR) systems by offering more fluid and intuitive user experiences \cite{K2View2025}.

However, deploying effective voice-based banking agents presents several challenges. First, financial interactions are often multi-turn, transactional, and stateful, requiring the agent to manage complex dialogues reliably. Second, financial applications must meet strict privacy and regulatory requirements such as GDPR and CCPA \cite{Finextra2025}, while also fostering user trust—an ongoing barrier to AI adoption in finance \cite{CFPB2024} \cite{RG_DebiasingStrategies2023}. Finally, the agent must integrate securely and seamlessly with backend banking systems to perform real-time operations like querying balances or executing transactions.

To address these challenges, this paper proposes a novel end-to-end system for secure and natural voice-based banking. The contributions of this work are as follows:
\begin{itemize}
    \item A hybrid system architecture that decouples telephony and media handling (via Twilio) from conversational logic (via Google Dialogflow CX), enabling scalable, channel-agnostic deployment
    \item  A controlled framework for tool-augmented large language models (LLMs), using Dialogflow CX’s state machine to reliably orchestrate calls to external banking APIs
    \item Implementation of advanced dialogue capabilities, including robust barge-in handling, seamless human handoff, and automated post-call summarization
    \item A multi-axis evaluation methodology covering functional task success, component-level performance (e.g., latency, ASR accuracy), and user-centric metrics such as perceived naturalness and trust.
\end{itemize}

\begin{figure*}
    \centering
    \includegraphics[width=0.8\linewidth]{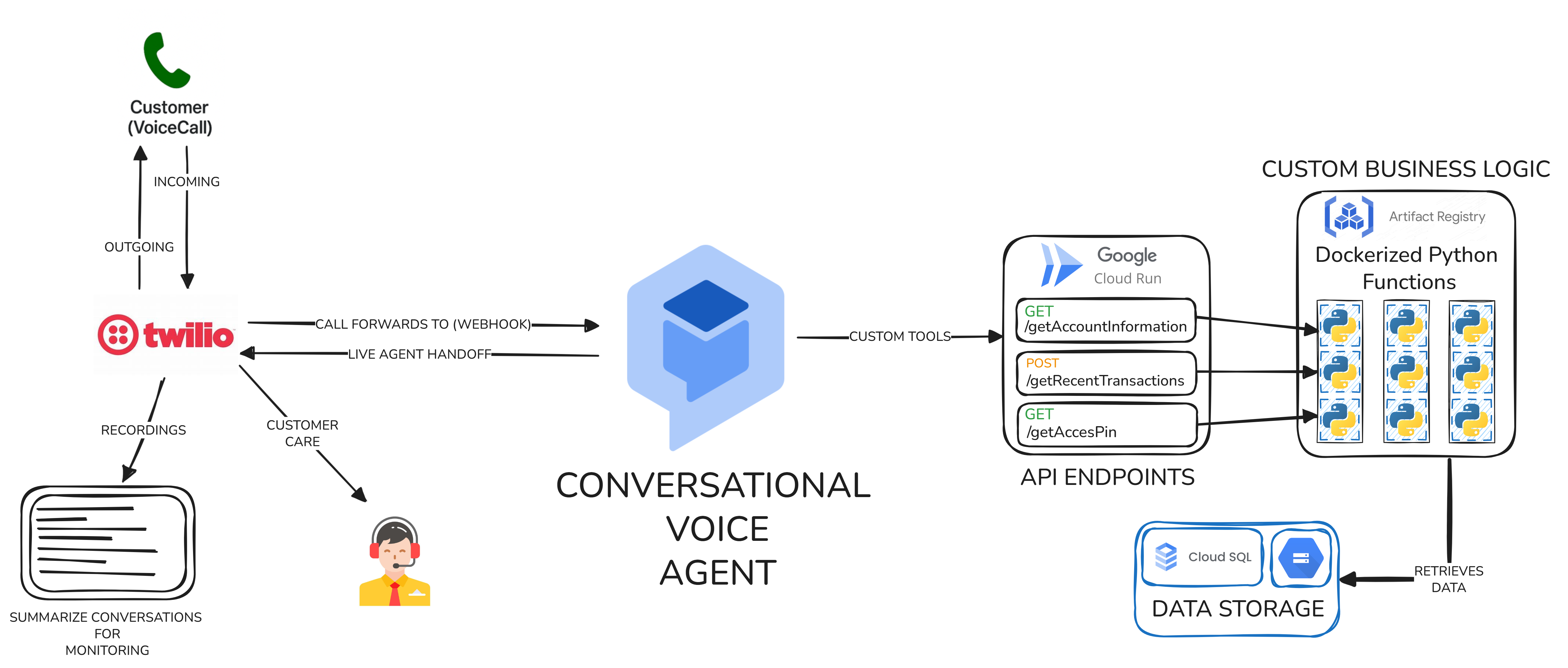}
    \caption{High-level Design of the Proposed Architecture}
    \label{fig:high-level}
\end{figure*}

\section{Literature Review}
Voice assistants are becoming more and more significant as access points to financial services, according to research on voice interfaces in financial services. Research shows that voice-based interfaces produce a more engaging and immersive consumer experience than text-based ones, thereby improving service perceptions and outcomes \cite{zierau2023voice}. This is particularly important for users who might have trouble using text-based interfaces because of literacy challenges, visual disabilities, or just a preference for spoken communication. Because financial data is sensitive, privacy and security issues are especially serious when it comes to banking chatbots. In their analysis of changing consumer expectations around privacy in AI banking systems, Birch and Rutter\cite{birch2023customers} point out a conflicting need for both privacy protection and personalization. According to their research, consumers are expecting banking interfaces to "know them" while upholding stringent data security.\\
PoornaPushkala et al. \cite{real-time-system-twilio-assembly} developed a robust real-time voice call system that uses Twilio and Assembly AI to transcribe and process customer queries. Their approach, by applying Natural Language processing, or NLP, and Naive Bayes classification, effectively demonstrated the extraction and automation of information, thus significantly reducing the workload of human support agents. However, this paper proposes that the system could be further enhanced by incorporating cloud storage integration and attaining greater autonomy. Lai et al. \cite{realtime-avatar} designed an avatar-based, tutor-like conversational AI to provide users with a more personalized interaction experience. However, the interface is intended mainly for desktop and PC users, and already presumes some expertise on the part of the users. Also, their work fails to discuss clearly how the system will scale up or integrate all the features proposed. On the other hand, the work by Mursi et al. \cite{finlingo} describes a multilingual solution to address banking needs in South Africa, outperforming traditional models. Despite its advantages, the remedy is burdened by the problem of constant internet access and user behavior. with the Finlingo app interface. Bhatia et al. \cite{bhatia2025ai} developed a multilingual chatbot that handles Hindi, English, and Hinglish effectively; however, it focuses on providing information and cannot conduct banking transactions. Bhatia et al.'s \cite{bhatia2025ai} and Oruganti's \cite{oruganti2020virtual} proposed that chatbots require internet connectivity on the customer's end.
 In response to the aforementioned limitations, our proposed architecture is built with the introduction of a cloud-native design, which provides a more scalable and autonomous solution. Our system operates with minimal human intervention and requires only human support when the customer explicitly requests it. Additionally, the proposed architecture addresses issues effectively: the telephony-based interface allows for use by non-technical users, while integration ensures scalable, high-performance support for cloud services. For multiple concurrent users without compromising the system's responsiveness.

\section{System Design and Architecture}
The overall architecture of the agent (Figure \ref{fig:high-level}) is envisioned to feature a modular, cloud-native design that incorporates Twilio for real-time telephony and a conversational voice agent based on large language models. Central banking operations are securely performed by containerized Python services on Google Cloud Run, accessing backend data through API endpoints linked to Cloud SQL. In addition to the automated conversation, the system also offers live agent handoff support for customer care services and post-call conversation summarization for monitoring, compliance, and quality assurance.

\begin{figure*}
    \centering
    \includegraphics[width=0.8\linewidth]{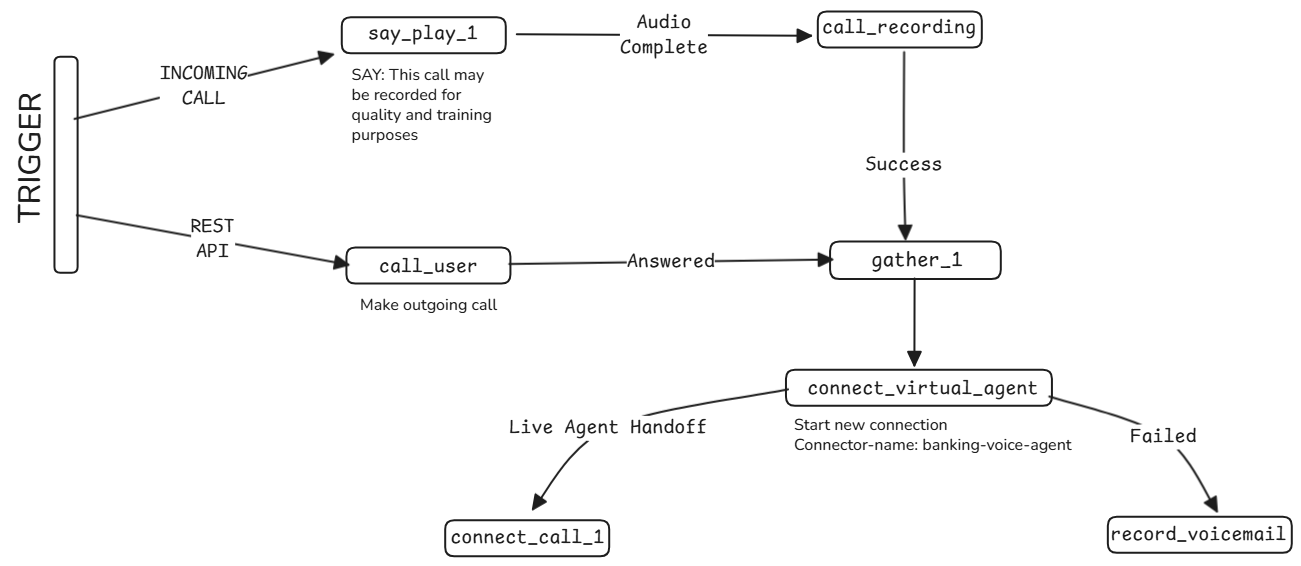}
    \caption{Twilio Studio Flow}
    \label{fig:twilio}
\end{figure*}

\subsection{User Interface}
We utilize Twilio \cite{official_twilio} as the telephony service. It allows the user to receive and make calls to the banking agent. This approach improves accessibility by allowing users without smartphones to interact with the system.

\subsection{Speech Recognition}
We use Google Cloud's Speech-to-Text service \cite{google-stt-phone} as part of the Dialogflow CX Conversational Agent platform. Specifically, we employ the \textit{phone-call} model, which is optimized for 8 kHz telephony audio, making it well-suited for real-time transcription over narrowband channels. During early prototyping, we evaluated open-source models such as Whisper \cite{whisper} and Vosk \cite{vosk-model}. However, these models exhibited lower transcription accuracy and higher latency in the telephony domain. Google’s STT service provided significantly better recognition performance for noisy or compressed voice signals commonly encountered in phone calls. According to Google documentation, the underlying STT engine leverages deep neural network architectures, including Conformer-based models, which are known for their accuracy in low-resource audio environments \cite{google-stt-conformer}.

\subsection{LLM Core with tools}
The Google Conversational Agent Platform introduces the concept of a playbook. Playbooks are a structured framework that defines how conversational agents respond to different user inputs and manage interactions. Playbooks allow for automating decision-making processes, creating dynamic conversation flows, and managing various tasks effectively. Task playbooks are the worker playbooks, in layman's terms, whereas Routine playbooks are the master playbooks. To elaborate:
\begin{enumerate}
    \item Task playbook: They are optimal for performing a single task, so good design practices encourage that each task playbook should be assigned a single task, for example, checking account balance in this case. Task playbooks can call each other if they require, but they can not directly call Routine playbooks. Task playbooks can pass output and input parameters between playbooks they want to, but can not add in the session parameters \cite{google-playbook-parameter}.
    \item Routine playbook: It manages multiple task playbooks under it. It can read and write session parameters that are accessible across the entire agent architecture. Routine playbooks have the privilege to call any playbook, whether another Routine or Task playbook, whereas Task playbooks can only call other Task playbooks but not any Routine Playbook. Routine playbooks also have access to the tools and flows, so they can use them directly if required.
\end{enumerate}
Google Conversational Agent platform defaults to using Gemini models as LLMs in the playbooks; these models serve as the core reasoning engine of the conversational agent. We use agent settings across all the playbooks; however, we have an option to customize LLM preference for each playbook individually. In this agent, we have used \texttt{\textbf{Gemini 2.5 Flash}} with an input token limit of 8000 tokens and an output token limit of 512 tokens, keeping the temperature at 0.9.

\subsection{Banking APIs}
Custom functions were developed to interface securely with the bank’s data source. All of these functions are developed individually to ensure scalability and modularity. Docker images were built for the functions alongside the environment, which are pushed to the Artifact Registry of Google Cloud. The containers are directly used by Cloud Run to deploy our custom functions to be accessed by our agent's LLM.

\subsection{Speech synthesis}
The voice agent uses the “Chirp” voice provided by Google’s Dialogflow CX platform for text-to-speech (TTS) synthesis \cite{google-tts-docs}. While Google does not publicly disclose the exact architecture behind this voice, it is part of their next-generation neural text-to-speech system designed for conversational quality. Based on public Google research papers, the Chirp voice likely utilizes Neural2, Tacotron 2, or neural codec-based models such as SoundStream \cite{google-stream} and AudioLM \cite{neural-voices}. These models are capable of producing expressive and natural-sounding responses suitable for real-time dialogue. The Chirp voice provides high intelligibility and prosody control, contributing to a smoother and more human-like conversational experience for telephony users.

\subsection{Authentication during conversation}
As a first step, our designed flow automatically fetches the user's number from the ongoing call. The phone number fetched is used to search in the bank's database for the registered phone number. This serves as the first authentication layer, verifying that only registered phone numbers can access the system. Furthermore, to perform critical tasks such as activating or deactivating the card, the user is required to enter a secret PIN, either by giving input from the dialpad or speaking it, which is associated with the phone number. The PIN needed for accessing telephony transactions is stored on the cloud server, mapped with the phone number and the account of the user. The agent uses a tool for verifying the PIN provided by the user. Only after the successful verification of the PIN, the user authorized to perform the task. This system ensures that the authorized owner of the account is performing all the transactions.

\begin{figure*}
    \centering
    \includegraphics[width=1\linewidth]{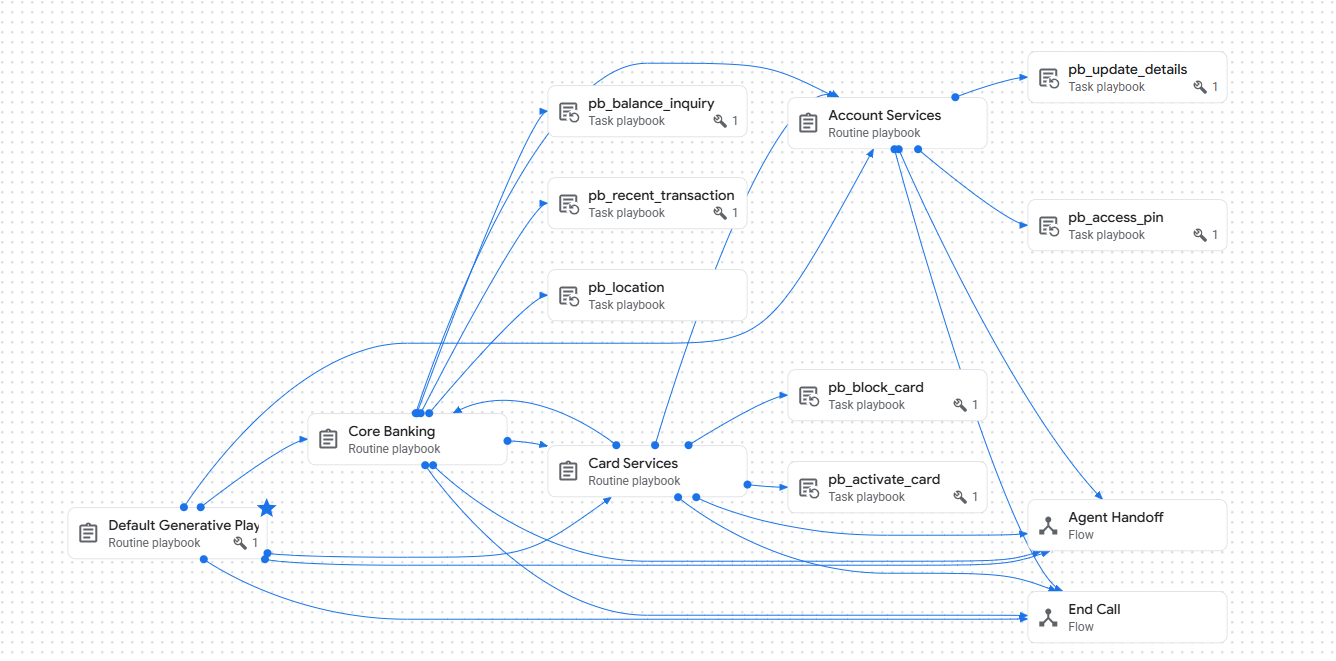}
    \caption{Conversational Voice Agent}
    \label{fig:google-agent}
\end{figure*}

\section{System Implementation}

The system was implemented as a modular and scalable cloud-native solution, integrating Twilio for real-time telephony and Dialogflow CX for voice-based conversational flows. Figures \ref{fig:twilio} and \ref{fig:google-agent} illustrate the detailed structure of the telephony and conversational agent workflows.

\subsection{Call Flow Design with Twilio Studio}

Figure \ref{fig:twilio} shows the Twilio Studio flow designed to manage incoming calls. The flow begins with a `Trigger` node that responds to incoming calls. The system then plays a message using `say\_play\_1` informing the user about call recording for quality and compliance. The recording says, "This call may be recorded for quality and training purposes. If you do not consent, hang up now." The `call\_user` widget initiates a call to the end user using the stored number (e.g., from a CRM or web form). After the call is answered, `call\_recording\_3` activates recording. The `gather\_1` node handles user input—either via DTMF tones (keypad input) or speech—and then connects the user to the virtual agent through the `connect\_virtual\_agent` node. This starts a session with the Dialogflow CX agent named `banking-voice-connection`. In case of connection failures or unresponsiveness, the flow includes logic for redirecting to voicemail (`record\_voicemail`) or human agent transfer via `connect\_call\_1`.

\subsection{Conversational Agent in Dialogflow CX}

As shown in Figure \ref{fig:google-agent}, the Dialogflow CX agent is structured into hierarchical playbooks and flows. The root agent contains multiple routine and task playbooks designed to manage distinct banking operations:

\begin{itemize}
    \item \textbf{Core Banking:} Handles balance inquiries, transaction history, and location services.
    \item \textbf{Card Services:} Manages activation and blocking of cards.
    \item \textbf{Account Services:} Enables users to update profile details or access PINs.
\end{itemize}

Each task is implemented as a separate playbook (e.g.,`pb\_balance\_inquiry`, `pb\_access\_pin`) to allow modular development and testing. The system supports a fallback generative model (`Default Generative Play`) for open-ended or out-of-domain user queries. Critical tasks are gated behind the `pb\_access\_pin` playbook, which verifies the user through PIN-based authentication before executing sensitive operations like card deactivation.

The agent also supports:
\begin{itemize}
    \item \textbf{Live Agent Handoff:} Seamless transfer to a human representative when needed. When the voice agent detects the intent like "I want to talk to customer care", "support", "connect me to a representative", etc., it triggers the live agent handoff flow, and the Google Conversational Agent platform gives the control back to the telephony provider, in this case, Twilio, with all the necessary output parameters. Then it is the responsibility of the telephony provider to handle the agent handoff configuration. Twilio provides a widget in their studio, which can directly connect the call to a static phone number; however, you can also enqueue a call if you wish to scale it to a call center environment. In the figure \ref{fig:twilio}, we can see that the "connect\_call\_1" widget handles connecting the call to a customer care's phone number when the "connect\_virtual\_agent\_1" widget requests a "Live Agent Handoff".
    \item \textbf{End Call Flow:} Graceful call termination with appropriate messaging.
\end{itemize}

\subsection{Session Management and Security}

Each session is initiated securely through Dialogflow CX using the built-in telephony connector. The user's phone number is automatically extracted and verified against the banking database. For sensitive functions, a second factor—user-entered PIN—is collected and verified before proceeding. All webhook interactions between Dialogflow and backend services are routed through HTTPS endpoints hosted on Google Cloud Run, with JWT-based authentication and audit logging.

\subsection{Backend Services and Containerization}

All banking operations (e.g., fetch balance, block card) are implemented as stateless Python microservices. Each microservice is containerized using Docker and deployed via Google Cloud Run. Container images are stored securely in Artifact Registry and follow CI/CD best practices.

\subsection{Logging and Post-Call Summarization}

Call sessions are logged using Google Cloud Logging and are monitored for latency, failure rates, and session durations. A post-call summarization service leverages Gemini to create human-readable summaries of the conversation, which are stored for compliance and support review.

\section{Evaluation and Results}
Standard evaluation metrics such as accuracy, precision, recall, and F1-score are commonly used for static classification tasks. However, conversational agents require multidimensional evaluation that reflects dynamic user interactions, system performance, robustness, and overall user experience. This section presents a comprehensive evaluation of the proposed banking voice agent across five categories: \textit{system performance}, \textit{conversational quality}, \textit{user experience}, \textit{security and compliance}, and \textit{robustness and stress testing}.

\subsection{System Performance}

\subsubsection{Latency}
From the user's spoken input to the system's audible response, the end-to-end latency was measured.  This covers the latency caused by Twilio's audio streaming, Dialogflow CX's intent detection, webhook fulfillment time, and text-to-speech synthesis.  Across several sessions, an average response latency of 1.12 seconds was recorded, staying within reasonable bounds for real-time conversational systems.  Calls to backend APIs for data retrieval were mainly linked to higher latency.

\subsubsection{Long-Duration Calls}
We used up to eight minutes of long-duration interactions to assess session persistence and stability.  The system demonstrated strong telephony integration and session handling between Twilio and Dialogflow CX by maintaining constant connection quality and session state over extended periods of time without degrading.

\subsubsection{Concurrent Call Handling}
Multiple concurrent calls were simulated in order to evaluate scalability.  We tested 5 concurrent call sessions that were supported by the system without observable performance deterioration or latency increase.  This demonstrates that the cloud-based deployment can efficiently handle multiple voice sessions. Furthermore, the system is robust enough to fairly more concurrent calls.

\subsubsection{Throughput and Uptime}
During continuous testing, the system showed a consistent throughput of multiple requests per minute with a 99.9\% uptime.  This attests to the architecture's dependability and preparedness for implementation on a large scale.

\subsubsection{Error Rate}
A steady backend integration between the conversational platform and telephone was demonstrated by the average rate of dropped calls or webhook timeouts, which was 0.1\%.

\subsection{Conversational Quality}

\subsubsection{Interruption Handling}
During natural conversations, users frequently talk over the system or interrupt.  We started user speech mid-response to test the system's capacity to manage such disruptions.  In 95\% of test cases, the agent was able to pause its output and process the new input, indicating flexible conversation and efficient real-time session management.

\subsubsection{Intent Recognition and Fallback Rate}
Intent detection accuracy was measured at 70\% for common banking intents, despite the fact that static classification accuracy is limited for conversational systems. 

\subsubsection{Unsupported Banking Requests}
Users asked for banking services that were not available (such as "apply for a new loan") in order to assess the agent's capacity to handle out-of-scope inquiries.  The agent ensured a consistent conversational experience by accurately identifying such cases and responding with fallback messages that directed users toward available services.

\subsubsection{Contextual Deviation}
In order to evaluate contextual drift handling, users purposefully started non-banking conversations.  Throughout the conversation, the system maintained contextual relevance by skillfully shifting the focus to banking-related subjects or offering a polite backup response.

\subsection{User Experience}

\subsubsection{Task Completion Rate}
The percentage of users who successfully finished particular tasks (such as recent transactions or balance inquiries) was known as the task completion rate, or TCR.  High task effectiveness was indicated by the observed TCR of 90\%.

\subsubsection{Average Interaction Time}
The average time to complete a banking task was 120 seconds, indicating that standard user requests were handled efficiently.

\subsubsection{User-Initiated Waits}
While completing a real-world task, users occasionally asked the agent to "wait."  A more human-like interaction flow was facilitated by the agent's ability to control brief pauses and sustain a conversational state for up to 15 seconds without session timeouts.

\subsection{Security and Compliance}

\subsubsection{Authentication Error Handling}
Intentionally entering incorrect authentication PINs was used to test security and user experience. In accordance with secure conversational design principles, the system appropriately denied access and terminated the call.

\subsubsection{Authentication Success Rate}
Consistent and secure user validation was demonstrated by the 100\% authentication success rate.

\subsection{Robustness and Stress Testing}

\subsubsection{Noise Robustness}
The accuracy of the system's speech recognition was assessed in a variety of background noise settings, such as an office with 40 dB of noise.  Under moderate noise, recognition accuracy decreased by only 5\%, demonstrating the underlying speech-to-text pipeline's resilience.

\subsubsection{Accent and Dialect Handling}
Users with varying linguistic backgrounds were used to test the agent. Across common English accents, intent recognition accuracy stayed above 80\%, indicating broad usability.

The thorough assessment shows that under a range of real-world conditions, the suggested voice-based banking agent maintains excellent performance, conversational robustness, and user satisfaction.  The system's feasibility for safe, scalable, and organic banking interactions via voice channels is validated by the combination of quantitative and qualitative evaluations.

\section{Conclusion and Future Work}
This paper presented a Google Conversational Agent-based architecture that efficiently serves all the requirements and purposes required for banking services while keeping in mind the security and authentication concerns. It has been tested on a wide variety of banking services such as balance inquiries, recent transactions, activation status of credit cards, and authentication-related tasks. It also allows for live agent handoff on complex or out-of-scope queries for smooth transitions to human agents when required. Various scenarios, such as long-duration calls, handling concurrent users, and dealing with interruptions or out-of-scope requests, are the basic ones on which it was tested. Similarly, it has been tested in noisy environments and with people having different accents and dialects to assess its robustness and to ensure the wide usability of the present prototype. These scenario variations can be used to show the ability of a voice agent to reliably, securely, and efficiently handle the demands of real-world user interactions and requests.\\
 All the data is securely stored in a cloud environment, which also ensures low response times. However, this system can be more secure by adding passive voice authentication, which eliminates the need for explicit security checks. Passive voice authentication works in the background and authenticates the user by the pitch and the tone of his voice, comparing with the already stored voice ID of the client. Additionally, OTP can be incorporated while performing critical tasks.

\bibliographystyle{IEEEtran}
\bibliography{references}
\end{document}